\documentclass[aps,prx,reprint,floatfix,superscriptaddress,nofootinbib,longbibliography]{revtex4-2}

\usepackage{amsmath,amssymb,bm,mathtools}
\usepackage{graphicx}
\usepackage{booktabs}
\usepackage{array}
\usepackage{xcolor}
\usepackage[colorlinks=true,linkcolor=black,citecolor=black,urlcolor=black]{hyperref}

\allowdisplaybreaks

\newcommand{\dd}{\mathrm{d}}
\newcommand{\ii}{\mathrm{i}}
\newcommand{\ee}{\mathrm{e}}
\newcommand{\sgn}{\operatorname{sgn}}

\newcommand{\tr}{\operatorname{tr}}

\newcommand{\calA}{\mathcal{A}}
\newcommand{\calM}{\mathcal{M}}
\newcommand{\calE}{\mathcal{E}}
\newcommand{\calB}{\mathcal{B}}
\newcommand{\bfq}{\bm q}
\newcommand{\bfr}{\bm r}
\newcommand{\bfk}{\bm k}
\newcommand{\bfu}{\bm u}

\newcommand{\bfcalA}{\bm{\mathcal{A}}}
\newcommand{\bfj}{\bm j}
\newcommand{\srcA}{\mathbb A}
\newcommand{\unitz}{\hat{\bm z}}
\newcommand{\vF}{v_{\!F}}

\newcommand{\nn}{\nonumber}
\long\def\rededit#1{{\color{red}#1}}

\begin{document}

\title{Finite-frequency anomaly-induced electromechanical response of Dirac fermions in deformed graphene}

\author{Ara Sedrakyan}
\affiliation{A.Alikhanyan National Science Laboratory, Yerevan 0036, Armenia}

\date{\today}

\begin{abstract}
A deformation of a graphene sheet changes more than the positions of the atoms.  In the low-energy Dirac theory it also produces geometric electron-phonon vertices.  One of these vertices acts as an emergent phonon gauge field, $\calA_\mu$, which couples to the same Dirac current as the electromagnetic vector potential.  This shared current vertex gives a direct route from mechanics to electronics: a moving deformation can generate a transverse electric current, and a deformation pattern with emergent phonon flux can bind electric charge.  We show that the coefficient of this mixed electromechanical response is the parity-odd current-current correlator of a massive Dirac cone.  For an insulating cone the coefficient is the one-cone Chern-Simons value, while for a doped cone in the local regime it is reduced by the Berry curvature factor $m/|\mu|$.  We apply the response to explicit deformations.  A traveling flexural wave generates a transverse second-harmonic current; a static ripple mixed with a dynamic phonon generates a transverse current at the drive frequency; and two non-collinear modes can generate charge modulation through the emergent phonon flux.  We keep the spin and valley sum explicit, so the paper shows when the one-cone anomaly becomes a charge current in graphene and when it instead appears in a valley, spin, or spin-valley channel.  For sublattice-gapped graphene with a valley-odd deformation gauge coupling, the two valleys add rather than cancel.  The experimentally sharp signature is a transverse electrical signal at twice the flexural-wave frequency, with a phase fixed by the sign of the sublattice gap and a gate dependence that crosses over from a gap plateau to a $1/|\mu|$ decay.  These direction, phase, frequency, and gate-voltage selection rules give clean tests of the anomaly-induced electromechanical channel in deformed graphene.
\end{abstract}

\maketitle

\section{Introduction}
\label{sec:intro}

Graphene is an unusually clean platform for studying how geometry acts on quantum matter.  Near the two inequivalent Dirac points, its low-energy quasiparticles are two-dimensional Dirac fermions.  A deformation of the sheet therefore does more than change the elastic energy.  It changes the local Dirac velocity, shifts the Dirac points, and can generate curvature-dependent scalar and mass-like terms.  This geometric view was developed in continuum treatments of fermions on curved two-dimensional surfaces and in low-energy theories of deformed honeycomb lattices \cite{KavalovKostovSedrakyan1986,SedrakyanStora1987,SinnerSedrakyanZiegler2011,SedrakyanSinnerZiegler2021,VozmedianoKatsnelsonGuinea2010,AmorimEtAl2016,deJuanSturlaVozmediano2012}.

Deformations of graphene are known to generate several electronic vertices, including scalar deformation potentials, velocity corrections, and valley-dependent pseudogauge fields \cite{SuzuuraAndo2002,Manes2007,CastroNetoEtAl2009,deJuanManesVozmediano2013}.  Strain engineering was proposed as a way to create large effective magnetic fields and pseudo-Landau levels \cite{GuineaKatsnelsonGeim2010,GuineaGeimKatsnelsonNovoselov2010,LowGuinea2010,MuchaFalko2012,GradinarMuchaSchomerusFalko2013}.  Experiments have observed large pseudomagnetic fields in graphene nanobubbles and in graphene on structured substrates, and optical measurements have shown that strain-induced pseudofields can strongly affect carrier dynamics \cite{LevyEtAl2010,JiangEtAl2017,KangEtAl2021}.  Recent work on periodic pseudogauge fields, elastic screening, and related two-dimensional Dirac systems further emphasizes that deformation-generated gauge fields are active electronic degrees of freedom rather than passive elastic parameters \cite{PhongMele2025,DeBeuleEtAl2025,HeidariParsiGhaemi2025}.

	A conceptually different approach to fermion--phonon interactions in graphene was developed in Ref.~\cite{SedrakyanSinnerZiegler2021}.  The main idea is geometric and uses the structure defined for manifolds.  A strongly deformed graphene sheet may not be described conveniently by a single flat coordinate chart.  Instead one covers the sheet by local patches/open sets and uses transition maps between overlapping patches,
\begin{equation}
  \hat\xi^{(a)}=f^{(ab)}[\hat\xi^{(b)}],
  \label{eq:patch_map_intro}
\end{equation}
which glue the local coordinate systems into one two-dimensional manifold; see Fig.~\ref{fig-1}.  This language is useful because local deformations, dislocations, and disclinations can then be treated within the same low-energy framework.

Starting from a deformed honeycomb lattice, let $\hat\mu_j$ with $j=1,2,3$ denote the three nearest-neighbor directions and let $t^j(\hat\xi)$ be the corresponding local hopping amplitudes.  The tight-binding Hamiltonian is
\begin{eqnarray}
  \label{H1}
  H &=& \sum_{j,\hat\xi} t^j
  \Big[
  \psi_A^+[\vec{X}(\hat\xi+\hat\mu_j)]
  \psi_B[\vec{X}(\hat\xi)]
  \nn \\
  &+&
  \psi_B^+[\vec{X}(\hat\xi)]
  \psi_A[\vec{X}(\hat\xi+\hat\mu_j)]
  \Big],
\end{eqnarray}
where $A$ and $B$ are the two sublattices and $\vec{X}(\hat\xi)$ is the position of the lattice site $\hat\xi$ in three-dimensional space.  Equation~\eqref{H1} simply says that every bond is allowed to know about the local shape of the sheet: the hopping strength and the bond direction can vary from point to point.  The continuum fields used later are the long-wavelength limit of this local bond geometry.

As shown in Ref.~\cite{SedrakyanSinnerZiegler2021}, local three-dimensional rotations of the hopping links can map the tangent frame of a deformed lattice onto a flat reference plane; see Fig.~\ref{fig-1}(b).  Under this local rotation the fermion fields transform as
\begin{equation}
  \Psi[\vec{X}(\hat\xi)] = \Omega[\vec{X}(\hat\xi)] \Psi'(\hat\xi).
  \label{eq:local_rotation_intro}
\end{equation}
Dropping the prime after the rotation, the Hamiltonian can be written as
\begin{eqnarray}
  \nn
  H &=& \frac{1}{2}\sum_{\hat\xi}\sum^{3}_{j=1}
  t^j
  {\Psi'}^+(\hat\xi)
  \Omega^+[\vec{X}(\hat\xi)]
  \sigma_1
  \\
  \label{H2}
  &\cdot&
  \Big[
  e^{- \overleftarrow{\partial}\cdot\hat\mu_j\sigma_3}
  +
  e^{\sigma_3\hat\mu_j\cdot\overrightarrow{\partial}}
  \Big]
  \Omega[\vec{X}(\hat\xi)]
  \Psi'(\hat\xi),
\end{eqnarray}
where the arrows show whether a derivative acts to the left or to the right, and $\sigma_{1,3}$ are Pauli matrices acting in the sublattice space.  This expression is not used below in its full lattice form.  Its role is to show the origin of the continuum vertices: local rotations and local hoppings generate a velocity vertex, a mass-like curvature vertex, and a gauge-like vector vertex.

\begin{figure}
	\centering
	\includegraphics[width=1.\linewidth]{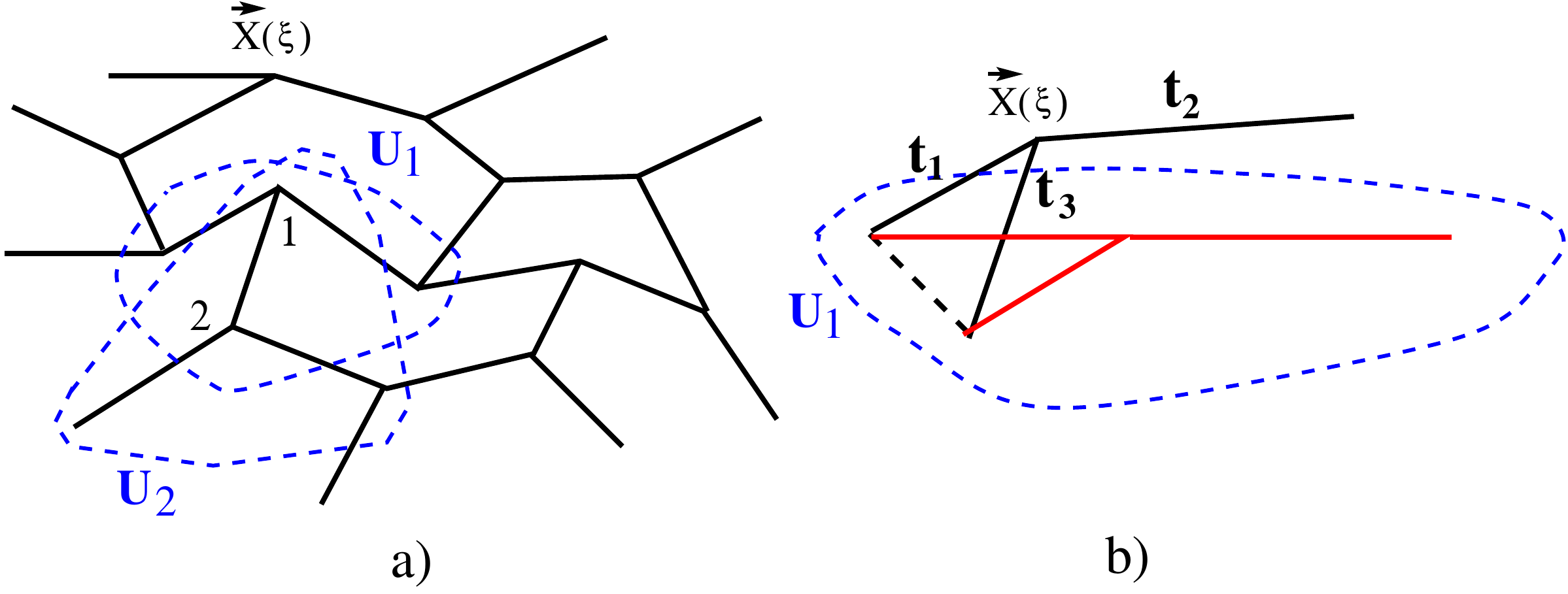}
	\caption{Local geometry of a deformed honeycomb sheet.  (a) A deformed lattice is described by local patches, here denoted $U_1$ and $U_2$, which cover neighboring lattice sites and the link between them.  The patch language keeps track of how the curved or distorted sheet is assembled from local coordinate systems.  (b) A site at position $\vec X(\hat\xi)$ in three-dimensional space is rotated into a local flat reference plane.  These local rotations are the microscopic source of the geometric gauge fields that appear in the continuum Dirac theory.}
		\label{fig-1}
\end{figure}

The same lattice Hamiltonian gives the local condition for a pair of Dirac nodes,
\begin{equation}
  \label{zero-point}
  \sum^{3}_{j=1} t^j e^{i \hat\mu_j\cdot \vec K} = 0,
\end{equation}
where the momentum $\vec K$ is the local position of a Dirac point in a given patch.  Geometrically, Eq.~\eqref{zero-point} means that the three complex hopping vectors must close into a triangle.  When this condition is satisfied, the deformation shifts the Dirac points but does not by itself destroy the low-energy Dirac quasiparticles.  After introducing the local two-dimensional metric, the low-energy kinetic term takes the form \cite{SedrakyanSinnerZiegler2021}
\begin{equation}
  \label{H5}
  H=\frac{i}{2}\int d{\hat\xi}~ \sqrt{g }\,
  \Psi^+(\hat\xi) \hat{\gamma}^\alpha
  \big(\overrightarrow{\partial}_\alpha-
  \overleftarrow{\partial}_\alpha\big)\Psi(\hat\xi),
\end{equation}
where $g=\det[g_{\alpha\beta}]$ and $\hat\gamma^\alpha$ are induced gamma matrices.  Equation~\eqref{H5} displays the metric part of the curved-surface Dirac operator.  The local rotations used to define the tangent frame also produce a non-Abelian $SU(2)$ connection.  In the weak-deformation limit, the component of this connection that couples as a vector source is the phonon gauge field $\calA_a$ used below.

In the present paper, we isolate the parity-odd part of the problem of fermion-phonon interactions via SU(2) gauge field.  In simple terms, this is the part of the response that changes sign when the Dirac mass or the orientation of a Dirac cone is reversed.  For one massive two-component Dirac fermion in $(2+1)$ dimensions, the same parity-odd determinant gives the Chern-Simons response associated with the parity anomaly \cite{RedlichPRL1984,RedlichPRD1984,Semenoff1984,Jackiw1984,Susskind1977,Haldane1988}.  Electron-phonon coupling can also generate Chern-Simons-like structures in Dirac systems \cite{SinnerZiegler2016,SinnerZiegler2019,BaskoAleiner2008}.  Here the key observation is very direct: if the fermion couples both to the electromagnetic vector potential $A_\mu=(A_0,\bm A)$ and to a deformation-induced vector field $\calA_\mu=(\calA_0,\bm{\calA})$, then the one-loop determinant contains a mixed electromagnetic-phonon Chern-Simons term.  The field $\calA_\mu$ is not a real electromagnetic field.  It is an emergent field that describes how the local deformation shifts and rotates the low-energy Dirac cone.
\begin{figure}[t]
	\centering
	\includegraphics[width=0.7\linewidth]{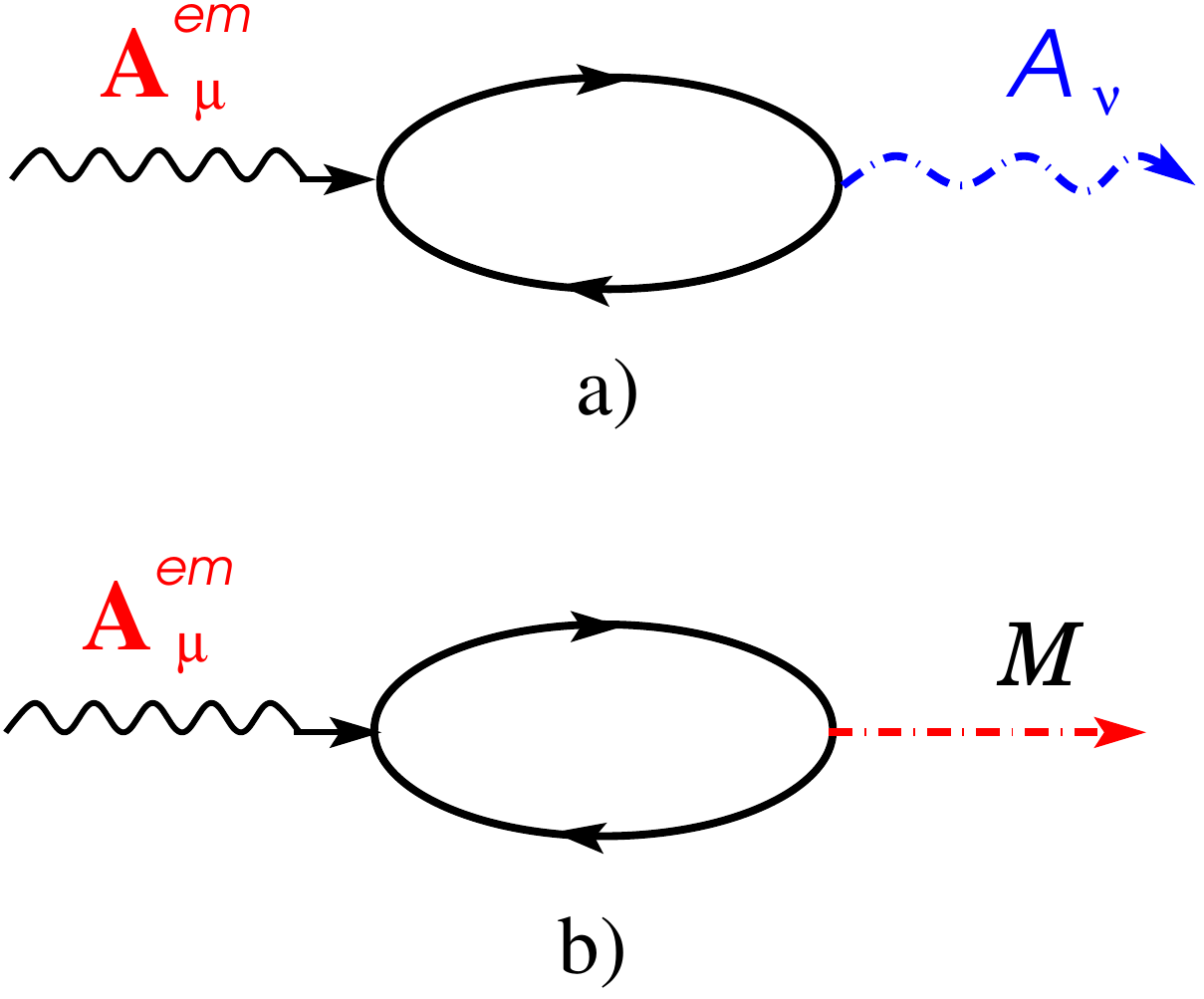}
	\caption{(Color online)
		The anomaly-induced signal is obtained from the mixed current-current bubble with one electromagnetic vertex, $A^{\rm em}_\mu$, and one phonon-gauge vertex, $\calA_\nu$.  Its parity-odd part is proportional to $\Pi^R_{\rm odd}\epsilon^{\mu\nu\rho}q_\rho$.  The lower diagram shows the ordinary mass-channel background from the scalar curvature vertex $\calM$; this background is useful for comparison but is not the mixed Chern-Simons response.  }
	\label{fig-1}
\end{figure}
\begin{figure}[h]
	\centering
	\includegraphics[width=1.02\linewidth]{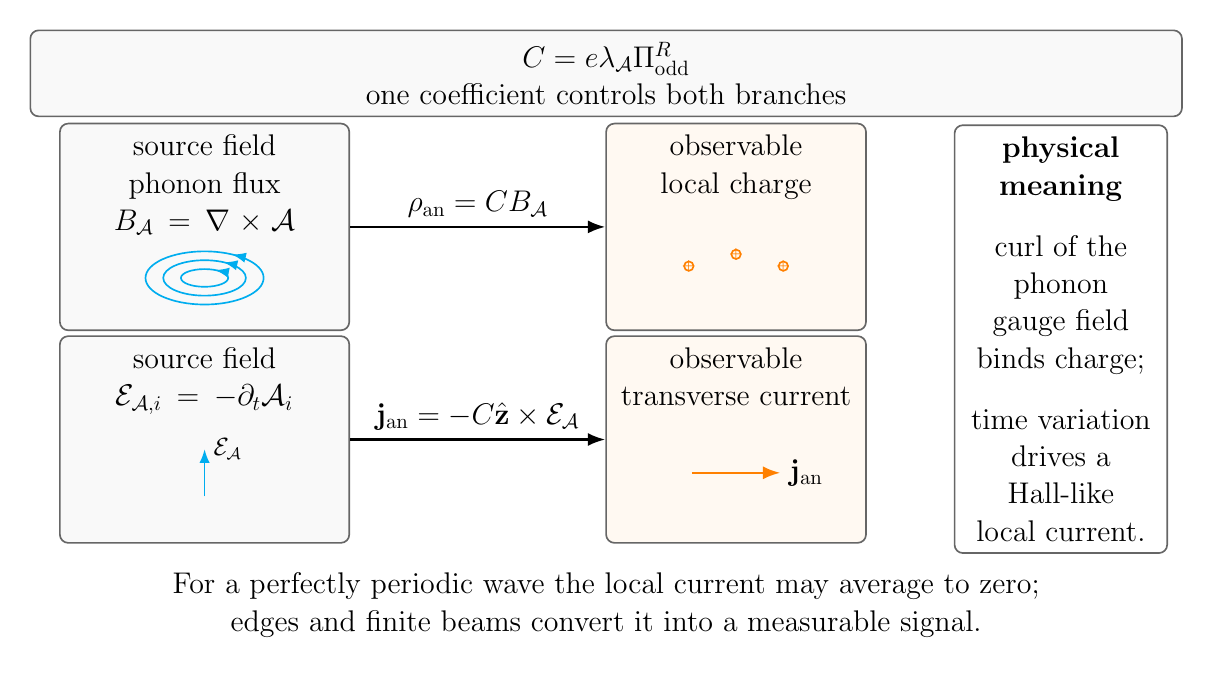}
	\vspace{-0.6cm}
	\caption{(Color online)
		 In real space the response has a simple meaning.  The phonon flux $\calB_{\calA}$ produces charge, while the phonon electric field $\bm\calE_{\calA}$ produces a transverse current.  The same coefficient $C=e\lambda_{\calA}\Pi^R_{\rm odd}$ controls both branches. }
	\label{fig-2}
\end{figure}

\begin{figure}[t]
	\centering
	\includegraphics[width=1\linewidth]{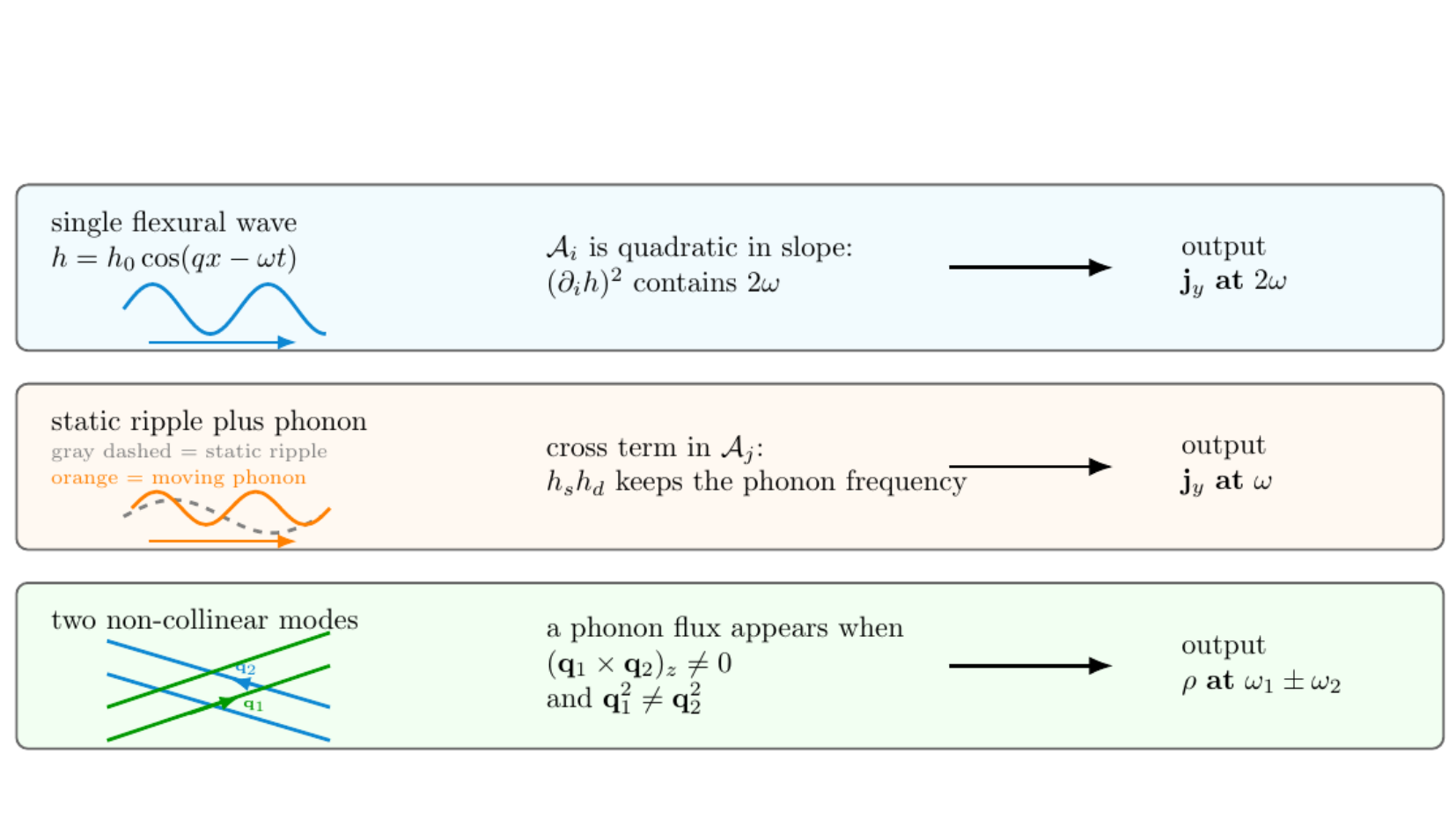}
	\caption{(Color online)
	 The nonlinearity of $\calA_i$ fixes the harmonic content: a single moving flexural wave gives a transverse current at $2\omega$, a static ripple plus a moving phonon gives a transverse current at the drive frequency $\omega$, and two non-collinear modes give charge at $\omega_1\pm\omega_2$ when their emergent phonon flux is nonzero.}
	\label{fig-3}
\end{figure}

The main results of the paper are the following.  First, for one spinless massive Dirac cone the deformation-induced electromagnetic current is\cite{footnote}
\begin{equation}
  j^\mu_{\rm an}(q)=
  e\lambda_{\calA}\Pi_{\rm odd}^R(q;\mu,m)
  \epsilon^{\mu\nu\rho}q_\nu\calA_\rho(q).
  \label{eq:central_intro}
\end{equation}
Here $q=(\omega,\bfq)$ is the external frequency-momentum, $m$ is the Dirac mass, $\lambda_{\calA}$ is the microscopic strength of the deformation gauge coupling, and $\Pi_{\rm odd}^R$ is the retarded parity-odd current-current correlator.  Equation~\eqref{eq:central_intro} is the central mixed electromechanical response.  It says that a time-dependent deformation drives a Hall-like current transverse to the phonon-induced electric field, while a deformation-induced flux binds charge.  The logic of the response and the associated Feynman diagram are summarized in Fig.~\ref{fig-1}-Fig.~\ref{fig-3}.

Second, the coefficient in Eq.~\eqref{eq:central_intro} is controlled by band topology only when the cone is insulating.  For $|\mu|\le |m|$ the long-wavelength coefficient is the one-cone Chern-Simons value, $-\sgn(m)/(4\pi)$.  For a doped cone, $|\mu|>|m|$, the controlled local coefficient is $-m/(4\pi|\mu|)$.  Thus doping does not destroy the Berry-curvature origin of the response, but it removes quantization because the Fermi surface participates.  This distinction is important in graphene, where finite-density response functions can contain nonanalytic structure from intraband excitations and Kohn anomalies \cite{ApresyanKhachatryanSedrakyan2015,Apresyan-Sedrakyan2019,SedrakyanMishchenkoRaikh2007,SedrakyanRaikh2008,SR08,WangRaikhSedrakyan2021Persistent,WangRaikhSedrakyan2021Interaction,WangSedrakyan2022Ballistic,ChaikaEtAl2025}.

Third, the geometric deformation gauge field is nonlinear in the displacement.  This gives clean phonon signatures.  A single traveling flexural wave produces a transverse current at $2\omega$, a static ripple mixed with a dynamic phonon produces a transverse current at $\omega$, and two non-collinear modes can produce charge modulation through the emergent phonon flux.  These frequency and geometry dependences are useful because they distinguish the parity-odd response from ordinary deformation-potential backgrounds.

Finally, graphene contains two valleys and two spins.  The one-cone result is therefore not, by itself, a charge-current prediction for graphene.  The valley chirality, the mass pattern, and the valley parity of $\lambda_{\calA}$ decide whether the response adds as a charge current or cancels as a charge current and survives in a valley, spin, or spin-valley channel.  This spin-valley bookkeeping is kept explicit throughout the paper.  It is closely tied to current studies of gapped Dirac systems, sublattice order, and valley-selective collective modes \cite{TarafdarSedrakyan2025,DuEtAl2025}.

This bookkeeping leads to a concrete experimental prediction for realistic graphene.  Consider sublattice-asymmetric graphene, for example graphene aligned to hBN or another substrate that distinguishes the two carbon sublattices.  Such a substrate produces a Semenoff-type gap: the mass has the same sign in the two valleys.  The ordinary strain pseudogauge vertex, by contrast, is valley odd \cite{SuzuuraAndo2002,Manes2007,deJuanManesVozmediano2013}.  The key point is that the two valleys also have opposite Dirac chirality.  The chirality sign and the pseudogauge sign therefore cancel each other, so the two valleys, and then the two spins, drive charge current in the same physical direction rather than canceling.  A moving flexural ripple should then produce a transverse electrical signal at exactly twice the mechanical drive frequency.  This second harmonic is not accidental: the emergent gauge field contains one factor of local slope and one factor of local curvature, so a sinusoidal ripple squares into a response at $2\omega$.  The signal has three sharp experimental fingerprints.  It is perpendicular to the ripple wave vector, its phase changes by $\pi$ when the sign of the sublattice mass is reversed, and a gate sweep changes its size in a predictable way: the amplitude is essentially on a plateau while the Fermi level lies in the gap and decreases inversely with $|\mu|$ once the Fermi level enters a band.  These features give a clean lock-in protocol for separating the anomaly-induced transverse current from ordinary longitudinal deformation-potential backgrounds.

 Importantly, an ordinary deformation potential can easily make a density signal at the mechanical drive frequency, but it does not naturally give a mass-odd, transverse, second-harmonic charge current with the Berry-curvature gate dependence described above.  This combination of signatures is what makes the proposed effect experimentally identifiable even when the microscopic coupling $\lambda_{\calA}$ is not known precisely.

The paper is organized as follows.  Section~\ref{sec:geometry} summarizes the deformed-graphene Dirac Hamiltonian and identifies the physical meaning of the phonon vertices.  Section~\ref{sec:response} defines the mixed electromagnetic-phonon response.  Section~\ref{sec:bubble} gives the one-loop parity-odd kernel in the controlled insulating, local, and optical limits.  Section~\ref{sec:current} converts the kernel into charge and current responses.  Section~\ref{sec:mass_channel} separates the conventional mass-channel background from the anomaly-induced current.  Section~\ref{sec:phonons} applies the formula to explicit phonon profiles.  Section~\ref{sec:valley} gives the spin and valley sum required for graphene.  Appendices contain derivational details; the last appendix derives the microscopic consistency checks needed to validate a tight-binding or numerical Kubo calculation.

\section{Dirac fermions on a deformed sheet}
\label{sec:geometry}

\subsection{Geometry of the sheet}

Let $\xi^a$ with $a=1,2$ be material coordinates on the sheet.  Latin indices $a,b=1,2$ label these material coordinates and are summed when repeated.  The embedding in three-dimensional Euclidean space is
\begin{equation}
  \bm X(\xi,t)=\xi^1\hat{\bm x}+\xi^2\hat{\bm y}
  +\bfu(\xi,t),
  \qquad
  \bfu=(u^1,u^2,h),
  \label{eq:Xdef}
\end{equation}
where $h\equiv u^3$ is the flexural displacement.  The induced metric and normal are
\begin{align}
  g_{ab}&=\partial_a\bm X\cdot\partial_b\bm X,
  \label{eq:metric}\\
  \bm n&=\frac{\partial_1\bm X\times\partial_2\bm X}
  {|\partial_1\bm X\times\partial_2\bm X|}.
  \label{eq:normal}
\end{align}
The continuum Dirac theory follows by expanding the tight-binding Hamiltonian around local Dirac points while rotating each local tangent frame to a reference plane.  The tangent vectors generate induced gamma matrices, and the local rotations generate a spin connection or, equivalently, a local rotation gauge field \cite{KavalovKostovSedrakyan1986,SedrakyanStora1987,SedrakyanSinnerZiegler2021}.

Only the weak-deformation form is needed below.  In units $\hbar=1$ and, unless stated otherwise, $\vF=1$, the one-valley Hamiltonian density can be written as
\begin{equation}
  \mathcal D[\bfu]=
  \ii\sigma_a\partial_a
  +\ii T_{ja}[\bfu]\sigma_j\partial_a
  +\sigma_a\calA_a[\bfu]
  +\sigma_3\calM[\bfu]+\cdots .
  \label{eq:Dphonon}
\end{equation}
Here $a=1,2$ is a spatial index on the sheet and $j=1,2,3$ labels Pauli matrices.  The three displayed vertices have distinct physical meanings.

First, $T_{ja}$ changes the local velocity matrix.  To leading order,
\begin{equation}
  T_{ja}=-\partial_a u^j+O(u^2).
  \label{eq:Tja}
\end{equation}
This is the emergent-metric or stress-tensor vertex.  It contributes to ordinary parity-even conductivity corrections.

Second, the mass-like field is controlled by curvature.  The leading flexural term is
\begin{equation}
  \calM[\bfu]=\frac{1}{2}\nabla^2 h
  +O(u\partial^2h,\partial u\partial h).
  \label{eq:Mleading}
\end{equation}
Here $\nabla^2=\partial_x^2+\partial_y^2$.  The field $\calM$ modulates the local Dirac gap and therefore the carrier density in a doped system.

Third, the spatial phonon gauge field begins at quadratic order in the deformation gradients.  A convenient leading expression is
\begin{equation}
\begin{split}
  \calA_a[\bfu]=-&\frac{1}{2}\Big[(\partial_a h)\nabla^2h \\
  &+(\nabla^2u_b)
  (\partial_a u_b+\partial_b u_a)\Big]
  +O(u^3).
\end{split}
  \label{eq:Aleading}
\end{equation}
The microscopic normalization depends on the hopping model and the definition of the continuum fields.  We therefore keep an explicit coupling $\lambda_{\calA}$ in all response formulas; physically, $\lambda_{\calA}\calA_a$ is the energy-scale vector vertex seen by the Dirac fermion.  Equation~\eqref{eq:Aleading} is the geometric, local-frame gauge field generated by the deformation.  It should not be confused with a real electromagnetic vector potential.  It is also more specific than the usual phenomenological strain pseudogauge field: both objects shift Dirac points and can be valley odd, but their microscopic prefactors and allowed higher-gradient terms need not be identical.  The physical distinction is useful.  A scalar deformation potential changes the local carrier density, a velocity vertex changes the metric seen by the Dirac particle, while $\calA_a$ acts as a vector source for the Dirac current.  Only the last vertex directly feeds the parity-odd current-current bubble used below.  For this reason the spin-valley sum in Sec.~\ref{sec:valley} is kept explicit.

This separation of vertices also clarifies the experimental strategy.  Scalar and velocity vertices are present in any deformed lattice and can create large conventional signals.  The anomaly channel is identified not by its absolute size alone but by its transverse direction, its odd dependence on the Dirac mass, and its spin-valley selection rules.

\subsection{Why the gauge vertex isolates the anomaly}

The velocity vertex $T_{ja}$ produces parity-even current responses.  Such terms are important for optical conductivity and for quantitative comparison with microscopic models, but they are not the most direct probe of the anomaly channel.  The mass vertex $\calM$ is also important, especially at finite density, because it gives a conventional density response proportional to $\partial n/\partial m$.

The gauge vertex $\calA_a$ is different.  It couples to the same Dirac current as an external vector potential.  Hence the mixed electromagnetic-phonon response is governed by the ordinary current-current correlator of a massive Dirac fermion.  Its parity-odd part gives the Chern-Simons term.  This observation makes the finite-frequency problem technically simple: once the one-loop tensor $\Pi^{\mu\nu}$ is known, the anomaly-induced electromechanical current follows by replacing one electromagnetic leg by the phonon gauge field.

\section{Response theory}
\label{sec:response}

\subsection{Single-cone action}

We first keep one two-component Dirac cone.  Spin and valley sums are restored in Sec.~\ref{sec:valley}.  The unperturbed real-time Hamiltonian is
\begin{equation}
  H_0(\bfk)=\vF(\sigma_x k_x+\sigma_y k_y)
  +m\sigma_z-\mu .
  \label{eq:H0}
\end{equation}
Here $\bfk$ is the momentum measured from the Dirac point, $\vF$ is the Fermi velocity, and the Pauli matrices $\sigma_i$ act on the sublattice pseudospin.  The mass $m$ opens a gap $2|m|$, while the chemical potential $\mu$ fixes the filling.
For the loop calculation it is useful to work in Euclidean space,
\begin{equation}
  S_0=\int_k \bar\psi_k
  [\gamma^\mu k_\mu+m]\psi_k,
  \quad
  k_\mu=(\omega_n+\ii\mu,\vF k_x,\vF k_y),
  \label{eq:SE}
\end{equation}
where $\psi$ and $\bar\psi$ are Euclidean Dirac fields, $\omega_n$ is a fermionic Matsubara frequency, and $\{\gamma^\mu,\gamma^\nu\}=2\delta^{\mu\nu}$.  We use
\begin{equation}
  \int_k\equiv T\sum_{\omega_n}
  \int\frac{\dd^2k}{(2\pi)^2}.
\end{equation}
Here $T$ in the Matsubara sum is temperature; it should not be confused with the deformation tensor $T_{ja}$.  We take $T=0$ in the final response formulas.  The external-momentum integral is denoted analogously by $\int_q$, where the bosonic external frequency is denoted by $\Omega_n$.  The sign of $m$ is physical because the parity-odd response changes sign under $m\to -m$.

The electromagnetic field and the phonon gauge field enter through the same current vertex.  This is the microscopic reason that an electromagnetic probe can detect the parity-odd phonon response.  We write
\begin{equation}
  \srcA_\mu=eA_\mu+\lambda_{\calA}\calA_\mu,
  \qquad
  \calA_0=0,
  \label{eq:Bdef}
\end{equation}
where $\srcA_\mu$ is only a combined source field, and $\calA_i$ is the deformation field in Eq.~\eqref{eq:Aleading}.  The curvature-induced mass modulation is included as
\begin{equation}
  m\rightarrow m+\lambda_M\calM[\bfu].
  \label{eq:Mcoupling}
\end{equation}
Here $\lambda_M$ is the microscopic coupling of the curvature-induced mass vertex.  With these definitions the Euclidean action to first order in external vertices is
\begin{equation}
\begin{split}
  S_E=S_0&+\int_{k,q}
  \bar\psi_{k+q}\gamma^\mu \srcA_\mu(q)\psi_k \\
  &+\lambda_M\int_{k,q}
  \bar\psi_{k+q}\calM(q)\psi_k+\cdots .
\end{split}
  \label{eq:fullSE}
\end{equation}

\subsection{Quadratic effective action}

Integrating out the fermions gives
\begin{equation}
  \ee^{-S_{\rm eff}}=\int D\bar\psi D\psi\,\ee^{-S_E}.
\end{equation}
To quadratic order,
\begin{equation}
\begin{split}
  S_{\rm eff}^{(2)}=&\frac{1}{2}\int_q
  \srcA_\mu(-q)\Pi^{\mu\nu}(q)\srcA_\nu(q) \\
  &+\lambda_M\int_q
  \srcA_\mu(-q)\Pi_M^\mu(q)\calM(q) \\
  &+\frac{\lambda_M^2}{2}\int_q
  \calM(-q)\Pi_{MM}(q)\calM(q)+\cdots .
\end{split}
  \label{eq:Seff2}
\end{equation}
The current-current bubble is
\begin{equation}
  \Pi^{\mu\nu}(q)=-\int_k
  \tr[\gamma^\mu G(k+q)\gamma^\nu G(k)],
  \label{eq:bubble}
\end{equation}
with
\begin{equation}
  G(k)=\frac{-\gamma^\mu k_\mu+m}{k^2+m^2}.
\end{equation}
The mixed scalar-current correlator is
\begin{equation}
  \Pi_M^\mu(q)=-\int_k
  \tr[\gamma^\mu G(k+q)G(k)].
  \label{eq:PiMdef}
\end{equation}
The scalar-scalar correlator $\Pi_{MM}(q)$ in Eq.~\eqref{eq:Seff2} is defined in the same way, with two mass vertices instead of current vertices.
The electromagnetic current four-vector $j^\mu_{\rm em}=(\rho_{\rm em},\bm j_{\rm em})$ induced by a deformation is obtained by differentiating with respect to $A_\mu$:
\begin{equation}
  j^\mu_{\rm em}(q)=
  e\lambda_{\calA}\Pi^{\mu\nu}(q)\calA_\nu(q)
  +e\lambda_M\Pi_M^\mu(q)\calM(q)+\cdots .
  \label{eq:jgeneral}
\end{equation}
The first term contains the anomaly channel because it is the current-current response with one electromagnetic leg and one deformation-gauge leg, as shown in Fig.~\ref{fig-1}(b).  The second is a conventional mass-channel background; it is allowed by symmetry but is not a Chern-Simons response.  This separation is central experimentally: the anomaly channel is transverse and odd in $m$, whereas the mass channel is mostly longitudinal or density-like and depends on ordinary compressibility.

The distinction is analogous to separating a Hall response from an ordinary longitudinal response.  Both can be generated by the same deformation, but only the parity-odd current-current bubble carries the Chern-Simons tensor $\epsilon^{\mu\nu\rho}q_\rho$.  This tensor is what rotates the phonon-induced field into a transverse electrical current.

\section{Parity-odd one-loop kernel}
\label{sec:bubble}

\subsection{Vacuum tensor}

At $\mu=0$ the Euclidean tensor decomposes as
\begin{equation}
\begin{split}
  \Pi_E^{\mu\nu}(q)=&
  (Q^2\delta^{\mu\nu}-q^\mu q^\nu)
  \Pi_{\rm even}^E(Q;m) \\
  &+\ii\epsilon^{\mu\nu\rho}q_\rho
  \Pi_{\rm odd}^E(Q;m),
\end{split}
  \label{eq:decompE}
\end{equation}
where $\Omega_n$ is a bosonic Matsubara frequency and
\begin{equation}
  Q=\sqrt{\Omega_n^2+\vF^2\bfq^2}
  \label{eq:QE}
\end{equation}
is the Euclidean length of the external energy-momentum.
The trace in Eq.~\eqref{eq:bubble} gives
\begin{equation}
\begin{split}
  N^{\mu\nu}=&2[k^\mu(k+q)^\nu+k^\nu(k+q)^\mu \\
  &-\delta^{\mu\nu}k\cdot(k+q)+\delta^{\mu\nu}m^2] \\
  &-2\ii m\epsilon^{\mu\nu\rho}q_\rho .
\end{split}
  \label{eq:trace}
\end{equation}
The last line is the parity-odd term.  After Feynman parametrization and shifting the loop momentum,
\begin{equation}
  \Delta(x)=m^2+x(1-x)Q^2 .
\end{equation}
The odd scalar coefficient is
\begin{equation}
  \Pi_{\rm odd}^E(Q;m)=
  -\frac{m}{4\pi}\int_0^1
  \frac{\dd x}{\sqrt{m^2+x(1-x)Q^2}} .
  \label{eq:oddint1}
\end{equation}
Therefore
\begin{equation}
  \Pi_{\rm odd}^E(Q;m)=
  -\frac{m}{2\pi Q}
  \arctan\frac{Q}{2|m|}.
  \label{eq:oddvac}
\end{equation}
for one two-component cone \cite{Apresyan-Sedrakyan2019,ApresyanKhachatryanSedrakyan2015}.  In the long-wavelength limit,
\begin{equation}
  \Pi_{\rm odd}^E(Q;m)=
  -\frac{\sgn m}{4\pi}
  +\frac{\sgn m}{48\pi}\frac{Q^2}{m^2}
  +O(Q^4/m^4).
  \label{eq:oddsmallQ}
\end{equation}
The first term is the Chern-Simons coefficient of an isolated massive Dirac cone.  It is the origin of the anomaly-induced part of the electromechanical response.

The even part is useful as a check on the normalization.  Gauge-invariant regularization gives
\begin{equation}
  \Pi_{\rm even}^E(Q;m)=\frac{1}{2\pi}
  \int_0^1\dd x\,
  \frac{x(1-x)}{
  \sqrt{m^2+x(1-x)Q^2}} .
  \label{eq:evenint}
\end{equation}
For $m=0$, Eq.~\eqref{eq:evenint} reduces to $1/(16Q)$, the standard scaling of a massless Dirac cone in two spatial dimensions.

\subsection{Finite chemical potential: controlled limits}
\label{subsec:finite_mu}

A chemical potential selects the rest frame of the Fermi sea.  Therefore, at finite density and finite momentum the retarded tensor is not a function only of the Lorentz scalar $Q_R^2=\vF^2\bfq^2-(\omega+\ii0^+)^2$.  The full clean-limit tensor can contain additional transverse structures built from the Fermi-sea velocity $u^\mu=(1,0,0)$, and intraband particle-hole processes can make the response nonanalytic.  This caution is familiar from finite-density Dirac response functions and from graphene magnetotransport and Kohn-anomaly physics \cite{ApresyanKhachatryanSedrakyan2015,SedrakyanMishchenkoRaikh2007,SedrakyanRaikh2008,SR08,WangRaikhSedrakyan2021Persistent,WangRaikhSedrakyan2021Interaction,WangSedrakyan2022Ballistic,ChaikaEtAl2025}.  The present work therefore uses only the limits in which the parity-odd coefficient is universal and unambiguous.

First, if the chemical potential lies in the gap, $|\mu|\le |m|$, no Fermi surface is present at $T=0$.  The finite-frequency kernel is the analytic continuation of the vacuum result,
\begin{equation}
  \Pi_{\rm odd}^R(\omega,\bfq;m)=
  -\frac{m}{2\pi Q_R}
  \arctan\frac{Q_R}{2|m|},
  \label{eq:odd_insulator}
\end{equation}
with
\begin{equation}
  Q_R=\sqrt{\vF^2\bfq^2-(\omega+\ii0^+)^2} .
  \label{eq:analytic_cont}
\end{equation}
The square root and arctangent are taken on the retarded branch.  This expression becomes complex when interband particle-hole production is allowed.

Second, in a doped cone, $|\mu|>|m|$, the local Hall coefficient is fixed by the Berry curvature of the occupied states.  Defining
\begin{equation}
  E_\mu=\max(|m|,|\mu|),
  \label{eq:Emu}
\end{equation}
one obtains
\begin{equation}
  \Pi_{\rm odd}^R(0,\bm 0;\mu,m)=
  -\frac{m}{4\pi E_\mu}.
  \label{eq:odd_local}
\end{equation}
Thus an insulating cone gives $-\sgn(m)/(4\pi)$, while a doped cone gives
\begin{equation}
  \Pi_{\rm odd}^R(0,\bm 0;\mu,m)=
  -\frac{m}{4\pi|\mu|},
  \qquad |\mu|>|m| .
  \label{eq:odd_doped_smallQ}
\end{equation}
This is one of the main controlled inputs of the paper.  It is the coefficient used for the doped long-wavelength phonon response, with corrections controlled by $\omega/E_\mu$ and $\vF|\bfq|/E_\mu$ away from particle-hole continua.  Physically, doping exposes only part of the Berry curvature of the massive Dirac band, so the Hall-like coefficient is reduced rather than quantized.

For reference, the homogeneous optical limit $\bfq=0$ is also elementary:
\begin{equation}
  \Pi_{\rm odd}^{R,{\rm opt}}(\omega;\mu,m)=
  -\frac{m}{4\pi(\omega+\ii0^+)}
  \ln\frac{2E_\mu+\omega+\ii0^+}
          {2E_\mu-\omega-\ii0^+}.
  \label{eq:odd_optical}
\end{equation}
It reduces to Eq.~\eqref{eq:odd_local} for $|\omega|\ll 2E_\mu$ and develops an imaginary part above the Pauli-blocked interband threshold $|\omega|=2E_\mu$.  Away from the insulating case, Eqs.~\eqref{eq:odd_local} and \eqref{eq:odd_optical} should not be replaced by a single $Q_R$-only formula at finite momentum.

\section{Anomaly-induced current}
\label{sec:current}

Keeping only the mixed parity-odd term gives
\begin{equation}
  S_{A\calA}^{\rm odd}=e\lambda_{\calA}\int_q
  A_\mu(-q)\,\ii\epsilon^{\mu\nu\rho}q_\rho
  \Pi_{\rm odd}(q)\calA_\nu(q).
  \label{eq:SAodd}
\end{equation}
Varying this action with respect to the electromagnetic source gives
\begin{equation}
  j^\mu_{\rm an}(q)=
  e\lambda_{\calA}\Pi_{\rm odd}^R(q;\mu,m)
  \epsilon^{\mu\nu\rho}q_\nu\calA_\rho(q).
  \label{eq:main_current}
\end{equation}
This is the main result of the paper.  It is a mixed Chern-Simons law: the electromagnetic current is generated by derivatives of the deformation gauge field.  The formula is local in the external field but nonlocal in the coefficient $\Pi_{\rm odd}^R(q)$, which contains the band-gap, finite-frequency, and finite-density physics.  The current is therefore transverse and changes sign when the parity-odd Dirac mass changes sign.  This sign reversal is a direct experimental diagnostic of the anomaly channel.

This is a response law for a gapped Dirac band, not as a force applied directly to the ions.  The moving lattice creates $\calA_\mu$; the massive Dirac fermions convert this source into an electrical current through their Berry curvature.  This is why the same deformation can give different observable channels after the spin and valley sum is taken.

In real time the response can be written in terms of the emergent phonon electric and magnetic fields,
\begin{align}
  \calB_{\calA}&=\partial_x\calA_y-\partial_y\calA_x,
  \label{eq:BAdef}\\
  \bm{\calE}_{\calA}&=-\partial_t\bfcalA-\bm\nabla\calA_0 .
  \label{eq:EAdef}
\end{align}
For one cone,
\begin{align}
  \rho_{\rm an}(q)&=e\lambda_{\calA}\Pi_{\rm odd}^R(q)
  \calB_{\calA}(q),
  \label{eq:rho_an}\\
  \bfj_{\rm an}(q)&=-e\lambda_{\calA}\Pi_{\rm odd}^R(q)
  \unitz\times\bm{\calE}_{\calA}(q).
  \label{eq:j_an}
\end{align}
Here $\unitz$ is the unit normal to the reference graphene plane.  These equations show the physical content of Eq.~\eqref{eq:main_current}: the phonon flux $\calB_{\calA}$ binds charge, while the phonon electric field drives a transverse current.
This convention gives the local continuity equation
\begin{equation}
  \partial_t\rho_{\rm an}+\bm\nabla\cdot\bfj_{\rm an}=0,
  \label{eq:continuity}
\end{equation}
because $\partial_t\calB_{\calA}+\partial_x\calE_{{\calA},y}-\partial_y\calE_{{\calA},x}=0$.  In the deformation geometries considered below, $\calA_0=0$.  Then
\begin{equation}
  \bfj_{\rm an}=e\lambda_{\calA}
  \Pi_{\rm odd}^R(\omega,\bfq)
  \unitz\times\partial_t\bfcalA.
  \label{eq:j_an_simple}
\end{equation}
This is the main result in its most transparent real-time form: a moving deformation produces a Hall-like current perpendicular to the time derivative of the emergent vector field (see Fig.\ref{fig-2}).
With the Fourier convention $f(\bfr,t)\propto\exp[\ii(\bfq\cdot\bfr-\omega t)]$,
\begin{equation}
  j_x=\ii\omega e\lambda_{\calA}\Pi_{\rm odd}^R\calA_y,
  \qquad
  j_y=-\ii\omega e\lambda_{\calA}\Pi_{\rm odd}^R\calA_x .
  \label{eq:j_components}
\end{equation}
The sign of a quoted real-space current depends on the convention for $\epsilon^{\mu\nu\rho}$ and for Fourier transforms.  The transverse structure, current conservation, and the value of the kernel are convention-independent.

For one insulating cone at long wavelength,
\begin{equation}
  \bfj_{\rm an}=-\frac{e\lambda_{\calA}}{4\pi}
  \sgn(m)\,\unitz\times\partial_t\bfcalA,
  \label{eq:j_quantized_onecone}
\end{equation}
up to the same convention sign.  For a doped cone,
\begin{equation}
  \bfj_{\rm an}=-\frac{e\lambda_{\calA}}{4\pi}
  \frac{m}{|\mu|}\,
  \unitz\times\partial_t\bfcalA,
  \quad |\mu|>|m|,
  \label{eq:j_doped}
\end{equation}
valid in the local regime $|\omega|,\vF|\bfq|\ll E_\mu$ and away from particle-hole continua.

The insulating and doped formulas therefore predict two simple regimes.  If the Fermi level lies in the gap, the response is controlled only by the sign of the mass.  If the Fermi level is moved into a band, the same Berry curvature is still present, but the occupied states no longer give a quantized coefficient; the amplitude decreases as $1/|\mu|$.

\section{Mass-channel background}
\label{sec:mass_channel}

The field $\calM[\bfu]$ changes the local Dirac mass.  At zero density and in the absence of a magnetic or emergent gauge flux, a scalar mass modulation alone does not generate a transverse charge current.  At finite density it changes the carrier density.  For one spinless cone,
\begin{equation}
  n(\mu,m)=\frac{\sgn\mu}{4\pi\vF^2}
  (\mu^2-m^2)\Theta(|\mu|-|m|).
  \label{eq:Dirac_density}
\end{equation}
Here $\Theta$ is the Heaviside step function.  Therefore, away from the band edge,
\begin{equation}
  \rho_M=e\lambda_M\frac{\partial n}{\partial m}\calM,
  \qquad
  \frac{\partial n}{\partial m}=
  -\frac{\sgn\mu\,m}{2\pi\vF^2}
  \Theta(|\mu|-|m|).
  \label{eq:rhoM}
\end{equation}
This density response is not topological: it is an ordinary compressibility effect caused by shifting the band edge.  It is nevertheless important experimentally because the leading curvature field $\calM\simeq\nabla^2h/2$ is linear in the flexural displacement.  In a measurement it can be larger than the anomaly signal, but it has different symmetry.  It does not by itself produce the transverse Hall-like current in Eq.~\eqref{eq:j_an}; instead it mainly gives density modulation at the same frequency as the curvature field.

This difference is useful in practice.  A compressibility signal follows the local curvature and is generally even under reversal of the Dirac mass.  The anomaly signal is transverse and odd in the mass.  Measuring direction, phase, and gate dependence therefore provides a practical way to separate the two channels.

The mass channel also modulates the anomaly coefficient itself:
\begin{equation}
  \Pi_{\rm odd}(m+\lambda_M\calM)=
  \Pi_{\rm odd}(m)+\lambda_M\calM
  \frac{\partial\Pi_{\rm odd}}{\partial m}+\cdots .
  \label{eq:PiOdd_mass_expand}
\end{equation}
Thus a dynamic curvature field can change an existing anomalous response generated by a static phonon gauge field.  This is a higher-order effect, but it provides a clean way to test the mass dependence of the parity-odd kernel.

\section{Explicit phonon profiles}
\label{sec:phonons}

\subsection{Single flexural wave: second harmonic}

Consider a traveling flexural wave,
\begin{equation}
  h(x,t)=h_0\cos(qx-\omega t),
  \qquad u_x=u_y=0 .
  \label{eq:flex_wave}
\end{equation}
With $\phi=qx-\omega t$, Eq.~\eqref{eq:Aleading} gives
\begin{equation}
  \calA_x=-\frac{1}{2}(\partial_xh)\partial_x^2h
  =-\frac{q^3h_0^2}{4}\sin(2\phi),
  \qquad
  \calA_y=0 .
  \label{eq:Aflex}
\end{equation}
The response is transverse.  The factor of two in frequency appears because the gauge field contains the product of the slope and the curvature of the wave.  Using Eq.~\eqref{eq:j_components} at the Fourier component $(2\omega,2q)$ gives
\begin{equation}
  j_y(x,t)=\frac{e\lambda_{\calA}}{2}
  \Pi_{\rm odd}^R(2\omega,2q)
  \omega q^3h_0^2\cos(2\phi).
  \label{eq:jy_flex}
\end{equation}
This is one of the main experimentally useful results of the paper.  The sign follows the convention of Eq.~\eqref{eq:j_an_simple}.  The robust prediction is the second-harmonic transverse signal with amplitude proportional to $\omega q^3h_0^2$.  The factor $q^3h_0^2$ is the product of a slope $qh_0$ and a curvature $q^2h_0$.  Thus the effect is strongest for fast, short-wavelength ripples while remaining within the small-slope regime.  For the realistic sublattice-gapped, valley-odd-coupling case derived in Sec.~\ref{sec:valley}, this local signal is a true charge current rather than only a valley current.  This makes the $2\omega$ transverse response a direct electrical signature of the anomaly channel.

In the one-cone insulating limit,
\begin{equation}
  j_y(x,t)\simeq
  -\frac{e\lambda_{\calA}}{8\pi}\sgn(m)
  \omega q^3h_0^2\cos(2\phi),
  \label{eq:jy_flex_quantized}
\end{equation}
again up to the global convention sign.  The measured charge response requires the valley sum discussed in Sec.~\ref{sec:valley}.

\subsection{Longitudinal in-plane wave}

For a longitudinal in-plane wave,
\begin{equation}
  u_x(x,t)=u_0\cos(qx-\omega t),
  \qquad u_y=h=0,
  \label{eq:long_wave}
\end{equation}
Eq.~\eqref{eq:Aleading} gives
\begin{equation}
  \calA_x=-\frac{1}{2}(\partial_x^2u_x)(2\partial_xu_x)
  =-\frac{q^3u_0^2}{2}\sin(2\phi),
  \qquad
  \calA_y=0 .
  \label{eq:Ainplane}
\end{equation}
Thus
\begin{equation}
  j_y(x,t)=e\lambda_{\calA}
  \Pi_{\rm odd}^R(2\omega,2q)
  \omega q^3u_0^2\cos(2\phi).
  \label{eq:jy_inplane}
\end{equation}
This has the same transverse and second-harmonic structure as the flexural result, with a different geometric prefactor.

The in-plane example shows that the second harmonic is not tied to bending alone.  It is a consequence of the quadratic geometric gauge field.  Whenever the relevant component of $\calA_a$ is built from two factors of a monochromatic deformation, the anomaly current appears at twice the drive frequency.

\subsection{Static ripple plus dynamic flexural phonon}

A single flexural wave gives a second-harmonic signal because $\calA_a$ is quadratic in $h$.  A linear-frequency signal is obtained by mixing a static ripple with a dynamic wave:
\begin{equation}
  h(x,t)=h_s\cos(Qx)+h_d\cos(qx-\omega t),
  \qquad h_d\ll h_s .
  \label{eq:static_dynamic}
\end{equation}
Keeping terms linear in $h_d$,
\begin{equation}
\begin{split}
  \calA_x^{(sd)}=-\frac{h_sh_d}{2}
  \Big[&Qq^2\sin(Qx)\cos(qx-\omega t) \\
  &+qQ^2\sin(qx-\omega t)\cos(Qx)\Big].
\end{split}
  \label{eq:A_static_dynamic}
\end{equation}
Equivalently,
\begin{equation}
\begin{aligned}
  \calA_x^{(sd)}=&-\frac{h_sh_d qQ}{4}
  \Big[(q+Q)\sin((q+Q)x-\omega t) \\
  &-(q-Q)\sin((q-Q)x-\omega t)\Big].
\end{aligned}
  \label{eq:A_static_dynamic_fourier}
\end{equation}
The Fourier components occur at momenta $q\pm Q$ and at the frequency $\omega$.  Hence
\begin{equation}
  j_y^{(sd)}=e\lambda_{\calA}
  \Pi_{\rm odd}^R(\omega,q\pm Q)
  \partial_t\calA_x^{(sd)} .
  \label{eq:jy_static_dynamic}
\end{equation}
This is another main result for experiment.  The geometry is attractive because the response is linear in the dynamic amplitude $h_d$; the static ripple supplies the second power of deformation required by the gauge vertex.  In practice this means that a patterned static deformation can convert the anomaly signal from the second harmonic to the fundamental phonon frequency.  The conversion is useful for detection because the electrical signal then appears at the externally applied phonon frequency, while its transverse direction and mass-odd sign still distinguish it from ordinary deformation-potential backgrounds.

\subsection{Two non-collinear modes: charge modulation}

The anomaly-induced charge density is proportional to the emergent phonon flux:
\begin{equation}
  \rho_{\rm an}=e\lambda_{\calA}\Pi_{\rm odd}^R\calB_{\calA},
  \qquad
  \calB_{\calA}=\partial_x\calA_y-\partial_y\calA_x .
  \label{eq:rho_flux}
\end{equation}
A single one-dimensional wave has zero flux.  A simple nonzero example uses two non-collinear flexural modes,
\begin{equation}
  h(\bfr,t)=h_1\cos\phi_1+h_2\cos\phi_2,
  \quad
  \phi_i=\bfq_i\cdot\bfr-\omega_i t .
  \label{eq:two_modes}
\end{equation}
Using $q_i=|\bfq_i|$, the cross term in the gauge field is
\begin{equation}
  \calA_a^{(12)}=-\frac{h_1h_2}{2}
  \left[q_{1a}q_2^2\sin\phi_1\cos\phi_2
  +q_{2a}q_1^2\sin\phi_2\cos\phi_1\right].
  \label{eq:A_two_modes}
\end{equation}
Taking the curl gives the explicit phonon flux
\begin{equation}
  \calB_{\calA}^{(12)}=
  \frac{h_1h_2}{2}(q_1^2-q_2^2)
  (\bfq_1\times\bfq_2)_z
  \sin\phi_1\sin\phi_2.
  \label{eq:B_two_modes}
\end{equation}
Equivalently,
\begin{equation}
\begin{aligned}
  \calB_{\calA}^{(12)}=&
  \frac{h_1h_2}{4}(q_1^2-q_2^2)
  (\bfq_1\times\bfq_2)_z \\
  &\times[\cos(\phi_1-\phi_2)-\cos(\phi_1+\phi_2)].
\end{aligned}
  \label{eq:B_two_modes_components}
\end{equation}
Thus the charge density contains sum and difference components,
\begin{equation}
  \rho_{\rm an}^{(12)}=e\lambda_{\calA}
  \Pi_{\rm odd}^R(\omega_1\pm\omega_2,
  \bfq_1\pm\bfq_2)\,
  \calB_{\calA}^{(12)} .
  \label{eq:rho_two_modes_explicit}
\end{equation}
This is one of the main geometric results of the paper.  It makes the conditions for charge modulation transparent.  The flux vanishes for collinear waves, and it also vanishes for two ideal plane waves with equal wavelength, $q_1=q_2$.  Nonzero charge modulation therefore requires an oriented two-dimensional deformation pattern together with unequal Laplacian eigenvalues or a more general spatial profile.

This condition gives a direct way to design or rule out charge-modulation experiments.  Merely crossing two equal-wavelength plane waves is not enough.  The deformation must produce a real emergent flux, in the same way that an electromagnetic magnetic field requires a vector potential with nonzero curl.

\section{Spin and valley structure}
\label{sec:valley}

Before the explicit spin-valley sum is taken, the response theory describes one spinless Dirac cone.  Graphene has two spin states and two valleys.  Let $\chi=\pm1$ label the valleys and choose
\begin{equation}
  H_\chi=\vF(\chi\sigma_x k_x+\sigma_y k_y)
  +m_\chi\sigma_z-\mu .
  \label{eq:Hvalley}
\end{equation}
The factor $\chi$ changes the orientation of the Dirac cone and therefore the sign of the parity-odd response.  If the phonon gauge coupling in valley $\chi$ is $\lambda_{\calA,\chi}$, the total anomaly coefficient is
\begin{equation}
  \Pi_{\rm odd}^{\rm tot}=N_s
  \sum_{\chi=\pm1}\chi\lambda_{\calA,\chi}
  \Pi_{\rm odd}^{(0)}(\omega,\bfq;\mu,m_\chi),
  \label{eq:valleysum}
\end{equation}
with spin degeneracy $N_s=2$.  The function $\Pi_{\rm odd}^{(0)}$ is the one-cone coefficient in Sec.~\ref{sec:bubble}, without the valley orientation or phonon-coupling factor.

This sum is not a technical detail; it decides the physical channel that an experiment can see.  The common cases are summarized in Table~\ref{tab:valley}.  We write $s=\pm1$ for the physical spin.  A valley-even deformation coupling has $\lambda_{\calA,\chi}=\lambda_{\calA}$, while a valley-odd, or pseudogauge, coupling has $\lambda_{\calA,\chi}=\chi\lambda_{\calA}$.  Ordinary time-reversal-symmetric strain gauge fields are usually valley odd.  If a microscopic convention absorbs this sign into the definition of $\calA_a$, Eq.~\eqref{eq:valleysum} is the invariant statement.

The experimentally sharp charge-current case is a Semenoff mass together with a valley-odd deformation gauge coupling.  This is the symmetry structure expected for sublattice-gapped graphene, including graphene on aligned hBN, driven by an ordinary strain pseudogauge field.  Put $m_{\chi s}=m$ and $\lambda_{\calA,\chi}=\chi\lambda_{\calA}$ in Eq.~\eqref{eq:valleysum}.  The two factors of $\chi$ cancel:
\begin{equation}
\begin{split}
  C_{\rm gr}^{\rm S}(\mu,m)&\equiv
  eN_s\sum_{\chi=\pm1}\chi\lambda_{\calA,\chi}
  \Pi_{\rm odd}^{(0)}(\mu,m) \\
  &=4e\lambda_{\calA}\Pi_{\rm odd}^{(0)}(\mu,m).
\end{split}
  \label{eq:Cgr_semenoff_deriv}
\end{equation}
Here $C_{\rm gr}^{\rm S}$ is the coefficient multiplying $\hat{\bm z}\times\partial_t\bm{\calA}$ in the graphene charge current, and the superscript ${\rm S}$ denotes the Semenoff mass pattern.  Using the controlled local coefficient in Sec.~\ref{subsec:finite_mu},
\begin{equation}
  C_{\rm gr}^{\rm S}=
  -\frac{e\lambda_{\calA}}{\pi}{\cal F}_{\mu m},
  \qquad
  {\cal F}_{\mu m}=\begin{cases}
  \sgn(m), & |\mu|\le |m|,\\[2pt]
  m/|\mu|, & |\mu|>|m| .
  \end{cases}
  \label{eq:Cgr_semenoff}
\end{equation}
This is a graphene-level result, not a one-cone statement.  It predicts a real charge response because the two valleys add after the spin sum.  For the flexural wave of Eq.~\eqref{eq:flex_wave}, Eq.~\eqref{eq:Cgr_semenoff} gives
\begin{equation}
  j_{y,{\rm gr}}^{\rm S}(x,t)=
  -\frac{e\lambda_{\calA}}{2\pi}
  {\cal F}_{\mu m}\,\omega q^3h_0^2
  \cos(2qx-2\omega t),
  \label{eq:jy_flex_graphene}
\end{equation}
up to the global sign convention stated below Eq.~\eqref{eq:j_components}.  Equation~\eqref{eq:jy_flex_graphene} is the sharp experimentally detectable prediction for realistic gapped graphene emphasized in this work.  It gives a transverse charge current at twice the mechanical drive frequency.  Its phase reverses when the sublattice mass reverses, while in the doped local regime its magnitude scales as $|m|/|\mu|$.  A gate sweep therefore provides a direct test: after the Fermi level leaves the gap, the signal decreases as $1/|\mu|$ while keeping the same geometric $\omega q^3h_0^2$ dependence.  Because the predicted current is transverse and occurs at a generated harmonic, it can be isolated by measuring the quadrature and phase of the $2\omega$ electrical response rather than by measuring a dc current.

This result is the clearest bridge from the single-cone anomaly to an electrical experiment in graphene.  The valley-odd coupling supplies one sign, the opposite valley chirality supplies the opposite sign, and their product is valley even.  Spin then simply doubles the charge signal.

\begin{table*}[t]
\caption{How the one-cone anomaly appears after the spin and valley sum in graphene.  The charge response survives only when the product of valley chirality, mass pattern, and deformation-coupling parity is even under the full spin-valley sum.  If this product is odd, the charge current cancels between valleys or spins, but a valley, spin, or spin-valley current can remain.}
\label{tab:valley}
\begin{ruledtabular}
\begin{tabular}{lll}
Mass/coupling pattern & Charge response & Main surviving channel if charge cancels \\
\hline
Semenoff mass $m_\chi=m$, valley-even $\calA$ & cancels & valley current \\
Semenoff mass $m_\chi=m$, valley-odd $\calA$ & adds & charge current \\
Haldane mass $m_\chi=\chi m_H$, valley-even $\calA$ & adds & charge current \\
Haldane mass $m_\chi=\chi m_H$, valley-odd $\calA$ & cancels & valley current \\
Kane-Mele mass $m_{\chi s}=\chi s m_{\rm SO}$, valley-even $\calA$ & cancels & spin Hall-type current \\
Kane-Mele mass $m_{\chi s}=\chi s m_{\rm SO}$, valley-odd $\calA$ & cancels & spin-valley current \\
Massless symmetric graphene & vanishes & none \\
Valley-polarized state & partial cancellation & mixed charge-valley current
\end{tabular}
\end{ruledtabular}
\end{table*}

There are therefore several experimentally distinct regimes.  The same one-cone anomaly can appear as different physical currents after the lattice symmetries are restored.  This point is particularly relevant for gapped graphene on aligned substrates and for valley-selective collective modes, where mass pattern and valley content are part of the observable physics \cite{TarafdarSedrakyan2025,DuEtAl2025}.  A substrate or interaction-driven order that generates a non-canceling mass pattern gives a charge current.  Valley-selective pumping can turn incomplete cancellation into a charge signal.  In pristine symmetry-balanced graphene the charge response cancels, but the same calculation predicts a valley current whose two valley components flow with opposite signs.

\section{Scaling and observability}
\label{sec:units}

In the formulas above, we set $\hbar=1$.  Restoring $\vF$ amounts to using
\begin{equation}
  Q_R=\sqrt{(\vF|\bfq|)^2-(\omega+\ii0^+)^2}.
  \label{eq:QR_units}
\end{equation}
If the Hamiltonian is written as
\begin{equation}
  H=\vF\sigma_i[-\ii\partial_i-eA_i]
  +\sigma_i\calA_i+m\sigma_z,
  \label{eq:Hunits}
\end{equation}
then $\calA_i/\vF$ has the same dimension as an electromagnetic vector potential.  In practice the safest procedure is to keep $\lambda_{\calA}$ explicit and fix it from the microscopic hopping model or from a tight-binding calibration.

For a flexural wave the anomaly-current scale is
\begin{equation}
  |j_y|\sim |e\lambda_{\calA}\Pi_{\rm odd}|
  \omega q^3h_0^2 .
  \label{eq:j_scale}
\end{equation}
The scaling is robust even when the microscopic prefactor is model dependent, because it follows from geometry and from the mixed Chern-Simons tensor structure.  The effect grows with frequency, slope $qh_0$, and curvature $q^2h_0$.  The continuum expansion requires
\begin{equation}
  qh_0\ll1,
  \qquad
  qu_0\ll1,
  \label{eq:small_def}
\end{equation}
and the low-energy Dirac description requires momenta well below the inverse lattice spacing.

A realistic measurement must separate the parity-odd signal from parity-even backgrounds.  Useful diagnostics are the transverse direction, the sign reversal under $m\to -m$, the $m/|\mu|$ dependence in the doped regime, and the harmonic content: the pure flexural response occurs at $2\omega$, while the ripple-assisted response occurs at $\omega$.  These signatures are insensitive to the nonuniversal value of $\lambda_{\calA}$, although the absolute magnitude is not.  High-mobility and ballistic graphene devices are therefore natural settings for testing the effect because the clean-limit transverse response is least obscured by disorder-induced relaxation \cite{WangSedrakyan2022Ballistic,ChaikaEtAl2025}.

For sublattice-gapped graphene with a valley-odd deformation gauge coupling, such as a graphene/hBN device in which the substrate opens a sublattice gap, the one-cone caveat is removed: the spin-valley sum gives the charge coefficient in Eq.~\eqref{eq:Cgr_semenoff}.  The experimentally relevant scale of the pure flexural signal is therefore
\begin{equation}
  |j_{y,{\rm gr}}^{\rm S}|=
  \frac{e|\lambda_{\calA}|}{2\pi}|{\cal F}_{\mu m}|
  \omega q^3h_0^2 .
  \label{eq:graphene_signal_scale}
\end{equation}
This expression contains three robust tests that do not require knowing $\lambda_{\calA}$ precisely.  First, the signal is transverse and appears at $2\omega$.  Second, its phase changes by $\pi$ if the sublattice mass changes sign.  Third, in the doped local regime,
\begin{equation}
  \frac{j_y(\mu_1)}{j_y(\mu_2)}=\frac{|\mu_2|}{|\mu_1|},
  \qquad |\mu_{1,2}|>|m|,
  \label{eq:gate_ratio}
\end{equation}
for fixed deformation and fixed mass.  This gate-dependence is a sharp way to separate the Berry-curvature anomaly channel from ordinary geometric backgrounds, which need not follow the same $1/|\mu|$ law.  In practice, one can drive a known ripple mode, detect the transverse electrical response with a lock-in amplifier at $2\omega$, and then repeat the measurement while sweeping the gate through the gap.  The expected pattern is overconstrained: the phase is fixed by the mass domain, the frequency by the deformation geometry, and the gate dependence by the Berry curvature of the massive Dirac bands.

A useful experimental sequence is therefore: first identify the generated harmonic, then rotate or redesign the deformation to confirm the transverse direction, and finally sweep the gate voltage.  The full anomaly signature is the combination of these tests, not just a single nonzero voltage.

The current in Eq.~\eqref{eq:j_an_simple} is a local conserved current.  For an infinite periodic traveling wave its spatial average over one period is zero; this is not a uniform dc transport current.  It can nevertheless be probed by local current or magnetic-field probes, by finite wave packets, by edge accumulation in a bounded sample, or by radiation at the generated harmonic.  The theory predicts the local source term and its symmetry; the conversion into a contact signal depends on boundary conditions and device geometry.

\section{Discussion and conclusions}
\label{sec:conclusion}

We have derived a finite-frequency electromechanical response generated by the parity-odd determinant of massive Dirac fermions in a deformed graphene sheet.  The key physical point is simple.  A deformation produces an emergent gauge field $\calA_\mu$ for the Dirac quasiparticles.  Because this field couples to the same current as the electromagnetic vector potential, the ordinary parity-odd current-current bubble immediately becomes a mixed electromagnetic-phonon response.

The main result is Eq.~\eqref{eq:main_current}.  It is a mixed Chern-Simons relation between the electromagnetic current and the deformation gauge field.  Equivalently, Eqs.~\eqref{eq:rho_an} and \eqref{eq:j_an} state that the emergent phonon flux binds charge and the emergent phonon electric field drives a transverse current.  In one insulating cone the coefficient is fixed by the sign of the mass.  In a doped cone the local coefficient becomes $-m/(4\pi|\mu|)$, so the signal decreases as the chemical potential moves away from the gap.  In the homogeneous optical limit the same coefficient becomes complex above the Pauli-blocked interband threshold, which gives a direct route to dissipative finite-frequency physics.

The phonon predictions are also main results.  A traveling flexural wave produces a second-harmonic transverse current proportional to $\omega q^3h_0^2$, Eq.~\eqref{eq:jy_flex}.  The factor $q^3h_0^2$ has a clear geometric origin: it is a slope times a curvature.  A static ripple mixed with a dynamic phonon produces a fundamental-frequency transverse current, Eq.~\eqref{eq:A_static_dynamic_fourier}, because the static ripple supplies one power of the deformation.  Two non-collinear modes generate charge modulation only when they produce a nonzero emergent phonon flux, Eq.~\eqref{eq:B_two_modes}; in particular, equal-wavelength ideal plane waves do not suffice.  These selection rules are useful experimental diagnostics because they separate the parity-odd response from ordinary longitudinal deformation-potential effects.

The final main point is the spin-valley sum.  The anomaly of a single cone is not automatically a charge anomaly in graphene.  The observed channel depends on the valley chirality, the mass pattern, and the valley parity of the deformation coupling.  For some mass and coupling patterns the response adds as a charge current.  For others it cancels in charge but survives as a valley, spin, or spin-valley current.  This cancellation structure is not a limitation of the theory; it is a sharp symmetry prediction.

The most direct graphene charge-current prediction occurs for a Semenoff mass and a valley-odd deformation gauge coupling.  This situation is relevant to sublattice-gapped graphene, including graphene/hBN devices, driven by ordinary strain pseudogauge fields.  In that case the valley chirality and the valley-odd coupling compensate each other, so the two valleys and two spins give the coefficient $C_{\rm gr}^{\rm S}=-(e\lambda_{\calA}/\pi){\cal F}_{\mu m}$.  A traveling flexural wave then generates the transverse charge current in Eq.~\eqref{eq:jy_flex_graphene}.  The experimentally sharp signatures are a $2\omega$ electrical response, a phase reversal with the sign of the sublattice mass, and a crossover from an insulating plateau to a $1/|\mu|$ gate dependence in the doped local regime.

The response derived here should be testable by looking for transverse harmonic currents, edge accumulation in finite samples, local magnetic fields generated by the circulating current, or radiation at the generated harmonic.  For the charge-adding graphene case, the cleanest protocol is a lock-in measurement at twice the flexural drive frequency, followed by a gate sweep through the gap and into the doped regime.  The experiment does not need to know the microscopic value of $\lambda_{\calA}$ in order to identify the effect: the robust signatures are the transverse direction, the sign change under $m\to -m$, the insulating plateau followed by the $m/|\mu|$ doped-regime decay, and the harmonic selection rules imposed by the deformation geometry.  Observing this full set of signatures would provide a direct electrical probe of the parity-odd electromechanical response of gapped graphene.

The broader message is that lattice geometry can probe the same Berry-curvature physics that is usually accessed by electromagnetic fields.  Deformed graphene therefore provides a mechanical handle on a topological response: by shaping and driving the sheet one can select the frequency, momentum, and spin-valley channel of the anomaly-induced current.

\section*{Acknowledgments}
The research was supported by Armenian HESC grants
21AG-1C024 and 24FP-1F039.

\appendix
\section{Derivation of the deformation vertices}
\label{app:vertices}

The induced gamma matrices on the deformed sheet are
\begin{equation}
  \hat\gamma^a=\partial_a\bm X\cdot\bm\sigma .
  \label{eq:gammahat}
\end{equation}
Expanding Eq.~\eqref{eq:Xdef} about a flat sheet gives
\begin{equation}
  \hat\gamma^a=\sigma_a+(\partial_a u^j)\sigma_j+\cdots .
  \label{eq:gamma_expand}
\end{equation}
After symmetrizing derivatives in the Hamiltonian, the leading correction has the form
\begin{equation}
  \ii T_{ja}\sigma_j\partial_a,
  \qquad
  T_{ja}=-\partial_a u^j+\cdots .
  \label{eq:T_appendix}
\end{equation}
The spin connection and curvature pieces generate the scalar and gauge vertices.  Keeping the terms required in the main text,
\begin{equation}
  \calM=\frac{1}{2}\nabla^2h
  +O(u\partial^2h,\partial u\partial h).
  \label{eq:M_appendix}
\end{equation}
\begin{equation}
\begin{split}
  \calA_a=&-\frac{1}{2}\Big[(\partial_a h)\nabla^2h \\
  &+(\nabla^2u_b)
  (\partial_a u_b+\partial_b u_a)\Big]+\cdots .
\end{split}
  \label{eq:A_appendix}
\end{equation}
The omitted terms are higher order in gradients and can be restored from the full geometric Hamiltonian when a specific deformation geometry requires them.

\section{Current-current bubble}
\label{app:bubble}

For $\mu=0$, the parity-odd contribution follows from the last term of Eq.~\eqref{eq:trace}:
\begin{equation}
\begin{split}
  \Pi_{\rm odd}^{\mu\nu}=&-2\ii m\epsilon^{\mu\nu\rho}q_\rho
  \int_0^1\dd x \\
  &\times\int\frac{\dd^3\ell}{(2\pi)^3}
  \frac{1}{(\ell^2+m^2+x(1-x)Q^2)^2} .
\end{split}
  \label{eq:odd_app_1}
\end{equation}
Using
\begin{equation}
  \int\frac{\dd^3\ell}{(2\pi)^3}
  \frac{1}{(\ell^2+\Delta)^2}
  =\frac{1}{8\pi\sqrt{\Delta}},
  \label{eq:int_standard}
\end{equation}
we obtain
\begin{equation}
  \Pi_{\rm odd}^{\mu\nu}=-\ii\epsilon^{\mu\nu\rho}q_\rho
  \frac{m}{4\pi}
  \int_0^1\frac{\dd x}{
  \sqrt{m^2+x(1-x)Q^2}} .
  \label{eq:odd_app_2}
\end{equation}
The remaining parameter integral is
\begin{equation}
  \int_0^1\frac{\dd x}{
  \sqrt{m^2+x(1-x)Q^2}}
  =\frac{2}{Q}\arctan\frac{Q}{2|m|},
  \label{eq:x_integral}
\end{equation}
which gives Eq.~\eqref{eq:oddvac} after matching to the tensor decomposition \eqref{eq:decompE}.

The even part is transverse in a gauge-invariant regularization.  Terms odd in the shifted loop momentum vanish, and the non-transverse part is removed by the same regularization that enforces the Ward identity,
\begin{equation}
  q_\mu\Pi^{\mu\nu}=0 .
  \label{eq:ward}
\end{equation}
This gives Eq.~\eqref{eq:evenint}.

\section{Finite-density local and optical limits}
\label{app:finite_density}

The local doped result can be derived without using Lorentz invariance.  For the massive Dirac Hamiltonian, the band energies are
\begin{equation}
  E_{s,\bfk}=sE_{\bfk},
  \qquad
  E_{\bfk}=\sqrt{\vF^2k^2+m^2},
  \qquad s=\pm1 .
  \label{eq:band_energies}
\end{equation}
The Berry curvature of band $s$ is
\begin{equation}
  \Omega_s(\bfk)=-s\frac{m\vF^2}{2E_{\bfk}^3}.
  \label{eq:berry_curvature}
\end{equation}
With our current convention, the intrinsic odd coefficient is the occupied Berry-curvature integral.  At $T=0$ this gives
\begin{equation}
  \Pi_{\rm odd}^R(0,\bm 0;\mu,m)
  =-\frac{m}{4\pi E_\mu},
  \qquad E_\mu=\max(|m|,|\mu|),
  \label{eq:berry_result_app}
\end{equation}
which is Eq.~\eqref{eq:odd_local}.

The homogeneous optical limit follows from the interband Kubo formula.  Pauli blocking sets the lower interband energy to $E_\mu$.  The odd coefficient is
\begin{equation}
  \Pi_{\rm odd}^{R,{\rm opt}}(\omega;\mu,m)=
  -\frac{m}{4\pi}\int_{E_\mu}^{\infty}
  \frac{\dd E}{E^2-(\omega+\ii0^+)^2/4} .
  \label{eq:opt_integral}
\end{equation}
Evaluating the integral gives Eq.~\eqref{eq:odd_optical}.  The logarithm shows explicitly how the real Hall coefficient at low frequency turns into a complex optical response above the threshold $|\omega|=2E_\mu$.

\section{Derivational consistency checks for a microscopic calculation}
\label{app:checks}

This appendix derives the checks that a microscopic tight-binding or numerical Kubo calculation must satisfy in order to reproduce the continuum response.  The derivations are included to make clear which parts of the result are universal and which part must be calibrated microscopically.  The only nonuniversal quantity is the value of $\lambda_{\calA}$; the tensor structure, the limiting coefficients, and the harmonic selection rules follow from gauge invariance, Berry curvature, and the geometric form of the deformation gauge field.

First consider gauge invariance.  On the lattice the electromagnetic current vertex is obtained by differentiating the inverse Green function with respect to the vector potential.  In the continuum limit this gives the Ward identity
\begin{equation}
  q_\mu\Gamma^\mu(k+q,k)=G^{-1}(k+q)-G^{-1}(k).
  \label{eq:ward_vertex_app}
\end{equation}
Multiplying Eq.~\eqref{eq:ward_vertex_app} into the current-current bubble gives
\begin{equation}
\begin{split}
  q_\mu\Pi^{\mu\nu}(q)
  =&-\int_k\tr\{[G^{-1}(k+q)-G^{-1}(k)] \\
  &\times G(k+q)\Gamma^\nu(k,k+q)G(k)\}.
\end{split}
  \label{eq:ward_bubble_app}
\end{equation}
The two terms reduce to the same integral after shifting the loop momentum, provided the regularization preserves gauge invariance.  Hence
\begin{equation}
  q_\mu\Pi^{\mu\nu}(q)=0 .
  \label{eq:ward_result_app}
\end{equation}
This identity is the microscopic origin of current conservation in Eq.~\eqref{eq:continuity}.  A numerical calculation that violates Eq.~\eqref{eq:ward_result_app} is not measuring the Chern-Simons response alone; it still contains gauge-noninvariant cutoff artifacts or inconsistent current vertices.

The second check is the anomaly coefficient of one insulating cone.  Taking $Q\to0$ in Eq.~\eqref{eq:oddint1} gives
\begin{equation}
  \Pi_{\rm odd}^E(0;m)
  =-\frac{m}{4\pi}\int_0^1\frac{\dd x}{|m|}
  =-\frac{\sgn(m)}{4\pi}.
  \label{eq:anomaly_limit_app}
\end{equation}
This half-quantized value is the one-cone parity anomaly.  A full honeycomb lattice has two valleys, so the physical charge response is obtained only after summing the valley orientations, masses, and deformation couplings as in Eq.~\eqref{eq:valleysum}.  Thus a microscopic calculation should not compare a two-valley lattice result with Eq.~\eqref{eq:anomaly_limit_app} directly.  It should compare each valley-resolved contribution with Eq.~\eqref{eq:anomaly_limit_app} and then perform the symmetry sum.

The third check is the doped local limit.  At finite density Lorentz invariance is absent, so a lattice calculation at finite $\omega$ and finite $\bfq$ should not collapse onto a single function of $Q_R$.  In the local Hall limit, however, the result is fixed by Berry curvature.  Using Eq.~\eqref{eq:berry_curvature}, the occupied-state integral gives
\begin{eqnarray}
  \Pi_{\rm odd}^R(0,\bm0;\mu,m)
 & =&\int\frac{\dd^2k}{(2\pi)^2}
  \sum_s n_F(E_{s,\bfk}-\mu)\,\Omega_s(\bfk)\nn \\
  &=&-\frac{m}{4\pi E_\mu}.
  \label{eq:doped_check_app}
\end{eqnarray}
For $|\mu|>|m|$ this becomes $-m/(4\pi|\mu|)$.  A microscopic calculation should approach this value in the order of limits appropriate to a local Hall coefficient, with $|\omega|$ and $\vF|\bfq|$ small compared with $E_\mu$ and away from particle-hole continua.

The fourth check concerns static deformations.  The anomaly-induced charge and current are
\begin{equation}
  \rho_{\rm an}=e\lambda_{\calA}\Pi_{\rm odd}^R\calB_{\calA},
  \qquad
  \bm j_{\rm an}=-e\lambda_{\calA}\Pi_{\rm odd}^R
  \hat{\bm z}\times\bm\calE_{\calA}.
  \label{eq:static_check_fields_app}
\end{equation}
A static pure-gauge deformation has $\bm\calE_{\calA}=0$ and $\calB_{\calA}=0$.  Equation~\eqref{eq:static_check_fields_app} then gives no anomaly-induced charge and no anomaly-induced current.  If a microscopic calculation finds a local equilibrium current in such a geometry, that current must be a magnetization, boundary, or parity-even background contribution rather than the mixed Chern-Simons transport current studied here.

The fifth check is the harmonic content of the deformation profiles.  For the flexural wave $h=h_0\cos(qx-\omega t)$, the gauge field is quadratic in the deformation,
\begin{equation}
  \calA_x=-\frac{1}{2}(\partial_xh)(\partial_x^2h)
  =-\frac{q^3h_0^2}{4}\sin[2(qx-\omega t)],
  \label{eq:flex_check_app}
\end{equation}
and therefore the transverse current is at $2\omega$.  In contrast, if $h=h_s\cos Qx+h_d\cos(qx-\omega t)$, the cross term proportional to $h_sh_d$ is linear in the dynamic amplitude and oscillates at $\omega$.  A microscopic calculation should reproduce this frequency conversion.  It is a direct consequence of the geometric form of $\calA_x$, not of a special property of the continuum approximation.

The sixth check is the phonon flux generated by two non-collinear modes.  Starting from Eq.~\eqref{eq:A_two_modes} and taking the curl gives
\begin{equation}
  \calB_{\calA}^{(12)}=
  \frac{h_1h_2}{2}(q_1^2-q_2^2)
  (\bfq_1\times\bfq_2)_z\sin\phi_1\sin\phi_2 .
  \label{eq:flux_check_app}
\end{equation}
Thus a two-dimensional orientation factor is necessary but not sufficient: the two ideal plane waves must also have unequal Laplacian eigenvalues, or the deformation profile must be more general than two equal-wavelength plane waves.  A numerical calculation should see no anomaly-induced charge density when this flux vanishes.

Together these derivations give a compact protocol for a microscopic test.  Compute the lattice current-current response with gauge-invariant current vertices, calibrate $\lambda_{\calA}$ from the deformation-induced Dirac-point shift, separate valley and spin contributions, and then verify Eqs.~\eqref{eq:ward_result_app}, \eqref{eq:anomaly_limit_app}, \eqref{eq:doped_check_app}, \eqref{eq:flex_check_app}, and \eqref{eq:flux_check_app}.  Passing these checks would establish that the lattice calculation is measuring the same anomaly-induced electromechanical response as the continuum theory.


\begin{thebibliography}{99}
\bibitem{footnote} We use $\hbar=1$.  Greek indices $\mu,\nu,\rho=0,1,2$ denote time and the two in-plane spatial directions; bold symbols such as $\bfq$ denote two-dimensional spatial vectors.  The tensor convention is $\epsilon^{012}=+1$.  The symbol $e>0$ denotes the magnitude of the electron charge.  The scalar $\mu$ denotes the chemical potential and should not be confused with the Greek spacetime index.  These definitions are stated here because they enter the main result immediately.
\end{thebibliography}
\end{document}